\documentclass[11pt,a4paper]{article}
 \usepackage{jheppub}

\skip\footins = 1\bigskipamount plus 2pt minus 4pt

\usepackage{amsfonts}
\usepackage{amssymb}
\usepackage{amsmath}
\usepackage[vcentermath]{youngtab}
\newcommand{\med}{\frac{1}{2}}
\newcommand{\proj}{\mathcal{P}}
\newcommand{\gen}{{\mathcal{J}}}

\newcommand{\be}{\begin{equation}}
\newcommand{\ee}{\end{equation}}
\newcommand{\bea}{\begin{eqnarray}}
\newcommand{\eea}{\end{eqnarray}}

\newcommand{\ID}{\mathbf{1}}
\newcommand{\CAS}{\mathrm{C}_2}
\newcommand{\Proba}{\mathrm{P}}

\newcommand{\CP}{\mathbb{C}\mathrm{P}}
\newcommand{\calL}{{\mathcal{L}}}

\newcommand{\DD}{\cdot\cdot}
\newcommand{\D}{D \!\!\!\! /~}
\newcommand{\ijtp}{\it{Int. J. Theor. Phys.}}
\newcommand{\hepth}{\tt{hep-th}}
\newcommand{\cmp}{\it{Commun.Math.Phys.}}  
\newcommand{\ijmpa}{\it{Int. J. Mod. Phys.}}
\newcommand{\npb}{\it{Nucl. Phys. B}}
\newcommand{\jhep}{\bf{JHEP}}
\newcommand{\jgp}{\it{J. Geom. Phys.}}
\newcommand{\plb}{\it{Phys. Lett. {\bf B}}}
\newcommand{\ap}{\it{Annals Phys.}}
\newcommand{\condmat}{\it cond-mat}

\newcommand\SU{\mathop{\rm SU}\nolimits}

\newcommand\Spin{\mathop{\rm Spin}\nolimits} 
\newcommand\Tr{\mathop{\rm Tr}\nolimits} 
 
\newcommand{\e}{\mathfrak{e}}
\title{Scalar and Spinor Field Actions on Fuzzy $S^4$: fuzzy $\CP^3$ as a $S^2_F$ bundle over $S^4_F$.}

\author[a]{ Julieta Medina}
\author[b]{ Idrish Huet}
\author[c]{ Denjoe O'Connor}
\author[c,d]{Brian P. Dolan}

\affiliation[a]{ Ciencias B\'asicas UPIITA-IPN, Av. IPN 2580, \\Col. La Laguna Ticom\'an, 07340 M\'exico, D.F. Mexico} 
\affiliation[b]{ Theoretisch-Physikalisches Institut, Friedrich-Schiller Universit\"at, \\
Jena, Max-Wien-Platz 1, D-07743, Th\"uringen, Germany.}
\affiliation[c]{School of Theoretical Physics, Dublin Institute for Advanced Studies\\
10 Burlington Road, Dublin 4, Ireland.}
\affiliation[d]{ Department of Mathematical Physics, \\
NUI Maynooth, Co. Kildare, Ireland}
\emailAdd{ jmedinag@ipn.mx} \emailAdd{idrish.huet@uni-jena.de}
\emailAdd{denjoe@stp.dias.ie}
\emailAdd{bdolan@thphys.nuim.ie}

\abstract{We present a manifestly $\Spin(5)$ invariant construction of
  squashed fuzzy $\mathbb{C}\mathrm{P}^3$ as a fuzzy $S^2$ bundle over
  fuzzy $S^4$.  We develop the necessary projectors and exhibit the
  squashing in terms of the radii of the $S^2$ and $S^4$. Our analysis
  allows us give both scalar and spinor fuzzy action functionals whose
  low lying modes are truncated versions of those of a commutative
  $S^4$.}

\begin{document}
\maketitle
\flushbottom
\tableofcontents

\section{Introduction}\label{section1}

Noncommutative spaces with four dimensions are an interesting way to
model space-time at small length scales. Amongst the simplest four
dimensional manifolds $S^4$ is, on account of the one-point
compactification of Euclidean field theories, an important model. We
focus on the fuzzy approach to noncommutative spaces, where the
function algebra is replaced by a sequence of finite dimensional
matrix algebras $\mathcal{A}_L$ and the metrical geometry is
determined, in our case, by a Laplacian acting on ``functions''
$\Delta$.  We will also present a Dirac type operator
that recovers the spectrum of the standard round Dirac operator 
on $S^4$ in a certain limit.

The fuzzy noncommutative 4-sphere, $S^4_F$, was first constructed in
\cite{GrPresKlim} but has been known for some time in different
settings \cite{Castelino}-\cite{Kimura} along with other 4-dimensional
fuzzy spaces \cite{Behr}-\cite{Castro:2005}. The key feature of
$S^4_F$ is that the algebra of functions does not form a closed
associative algebra. This can be understood since the product of two
``functions'' takes one out of the algebra of functions of $S^4_F$ and
a projection is necessary to bring the product back
\cite{Ramgoolam_0105006}.

Here we follow the line presented in \cite{Medina_OConnor} where the
algebra is associative but it includes modes which do not belong to
the fuzzy 4-sphere. The quantized version of $S^4$ can be constructed
only in an indirect manner if one demands associativity of the
algebra, this is a consequence of the fact that $S^4$ does not admit a
Poisson structure. The approach taken here is based on the fuzzy
complex projective spaces, first given in \cite{starprod_CPN} and
further explored in \cite{Dolan:2007}-\cite{Huet:2011}, and in the
fact that $\CP^3$ is a fibration over $S^4$. In this context a
construction for the scalar theory on a fuzzy 4-sphere was first
carried out as a Hopf fibration in \cite{GrPresKlim}, but without a
method of suppressing the unwanted modes. The necessary suppression mechanism
was supplied in \cite{Medina_OConnor}.

A method to obtain an effective scalar field theory on $S^4_F$ was
given in \cite{Medina_OConnor}, there, an algebraic approach was taken
to eliminate the unwanted modes by constructing a positive definite
operator whose kernel consists of exactly all the modes in $\CP^3_F$
that belong to $S^4_F$, this operator was interpreted as a
modification of the Laplacian. In the present work we give a
geometrical interpretation of the suppresion mechanism in terms of the
fibre bundle picture for $\CP^3$.

In section \ref{section2} we present a brief review of the aspects
needed of $\CP^N_F$ and $S^4_F$, we follow essentially
\cite{starprod_CPN}, \cite{Medina_OConnor}.  Section \ref{section3}
presents the construction of our case of interest, $\CP^3$, first as a
$\Spin(6)$ and then as a $\Spin(5)$ adjoint orbit. It continues with
the calculation of the invariant line element and isotropy subgroup in
both approaches using the Maurer-Cartan forms of the aforementioned
groups, this is done only at a particular fiducial point that we call
the ``north pole'', by equivariance this suffices. Section
\ref{section4} presents a one-parameter dependent squashed Laplacian
$\Delta_h$ which fixes the symmetry of $\CP^3_F$ to be $\Spin(5)$
instead of the ``round'' $\Spin(6)$ symmetry.  This Laplacian turns
out to be an interpolation of $\Spin(5)$ and $\Spin(6)$ quadratic
Casimir operators. Section 5 deals with the use of the $*$-product map
to construct the commutative analogue of $\Delta_h$. The metric of the
squashed $\CP^3$ is obtained from the squashed Laplacian explicitely
as a combination of projectors. The line element of the bundle $\CP^3
\rightarrow S^4$ is computed and reinterpeted in terms of the found
radii of the fibre and base space. In section \ref{section6}, in the
spirit of \cite{Dolan_OConnor}, we present a first order operator on
$\CP^3_F$ that projects down to the Dirac operator in a certain limit
and hence give a prescription to construct an action for fermions
supressing the unwanted degrees of freedom. Section \ref{section7}
presents our conclusions.

\section{Review of $\CP^N_F$ and $S^4_F$}\label{section2}

In the usual construction, $\CP^N$ is defined as the space of all 
equivalence classes $[\psi]$ of unit vectors 
$\psi \in \mathbb{C}^{N+1}$, $|\psi| =1$, given by the equivalence 
relation: $\psi_1 \sim \psi_2$ if and only if 
$\psi_1 = e^{\imath \varphi} \psi_2$ for some $\varphi \in (0,2\pi]$. 
We follow closely the presentation in \cite{starprod_CPN} where the 
general details are given, and specialize later to the case under 
study of $\CP^3$. It was shown in \cite{starprod_CPN} that each 
equivalence class is associated with a hermitian rank 
one projector in $\mathbb{C}^{N+1}$, 
$\mathcal{P} = \psi \otimes \psi^{\dagger}$, 
we have then the following alternative definition of $\CP^N$

\be
\CP^N : = \{ \proj \in Mat_{N+1} : \proj^2 = \proj = \proj^{\dagger}, ~ \Tr \proj = 1 \} \label{cpndef}.
\ee

Each projector $\proj$ is associated with a point in $\CP^N$, a
coordinate system is introduced by expanding the projector in the
basis of matrices given by the identity and the generators of
$su(N+1)$ in the fundamental representation, denoted by $\{
\Lambda_\mu, ~~\mu=1,\ldots, N^2 + 2N \}$:

\begin{equation}
\proj = \frac{{\bf 1}}{N+1} + \frac{1}{\sqrt{2}} \xi_\mu \Lambda_\mu \label{proj}.
\end{equation}
The generators have been chosen to be orthogonal and with such
normalization that their algebra is

\be
\Lambda_\alpha \Lambda_\beta = \frac{2}{N+1} \delta_{\alpha \beta}{\bf 1} + \med (d_{\alpha \beta \gamma} + \imath f_{\alpha \beta \gamma}) \Lambda_\gamma \label{algebra}.
\ee
The conditions in (\ref{cpndef}) together with (\ref{proj}) and
(\ref{algebra}) result into a set of quadratic constraints for the
real coordinates $\xi_\mu$,

\be \label{constraints}
\xi_\mu \xi_\mu = \frac{N}{N+1},\quad  d_{\alpha \beta \gamma} \xi_\alpha \xi_\beta = \sqrt{8} \left(\frac{N-1}{N+1} \right)\xi_\gamma,
\ee
these constraints describe the embedding $\CP^N \hookrightarrow
\mathbb{R}^{N^2 + 2N}$, wherefrom the coordinates $\xi_\mu$ can be
seen to be a globally well defined overcomplete coordinate system. The
metric ${\bf P}$, complex structure $\bf{J}$, and K\"ahler structure
$\bf{K}$ on $\CP^N$ were found in \cite{starprod_CPN} to be given as

\bea \nonumber
\bf{P}_{\alpha \beta} &=& \frac{2}{N+1} \delta_{\alpha \beta} + \frac{1}{\sqrt{2}} d_{\alpha \beta \gamma} \xi_\gamma  - 2 \xi_\alpha \xi_\beta, \\ \label{geometrical}
\bf{J}_{\alpha \beta} &=& \frac{1}{\sqrt{2}} f_{\alpha \beta \gamma} \xi_\gamma, \\ \nonumber
   \bf{K}   &=& \med (\bf{P}+\imath \bf{J}).
\eea
Notice that the complex structure satisfies ${\bf J}^2 = -\bf{P}$.

One may obtain the fuzzy complex projective space $\CP^N_F$ by
considering the algebra of functions to be the full matrix algebra
given as

\be \label{matalgebra} Mat_{d_L^N} = \underbrace{\Yboxdim{12pt}
  \young(\ \DD\ )}_{L}\otimes\underbrace{ \overline{\Yboxdim{12pt}
    \young(\ \DD\ )}}_{L} \ee 
whose decomposition into irreducible representations of $\SU(N+1)$
corresponds with the expansion into polarization tensors of a function
on $\CP^N_F$. The dimension of the matrix algebra (\ref{matalgebra})
is $d_L^N = \binom{L+N}{N}$. The right-invariant vector fields induced
by the action of $\SU(N+1)$ are $\mathcal{L}_\mu =
\frac{\imath}{\sqrt{2}} J_{\mu \nu}\frac{\partial}{\partial \xi_\nu}$,
and in the fuzzy realization they take the form $ad(L_\mu)$ where
$L_{\mu}$ are the generators of the totally symmetric irreducible
representation; the associated Laplacian is then the quadratic Casimir
operator $\Delta = \frac{1}{R^2} (ad(L_\mu))^2 $, and reflects the
$\SU(N+1)$, hereafter called ``round'', symmetry of $\CP^N_F$. The
parameter $R$ is a length scale that fixes the size of $\CP^N$.  We
will analize in what follows a deformation of the Laplacian which
breaks the round symmetry and corresponds to a Kaluza-Klein-type
\cite{Salam} fuzzy space, first constructed in \cite{Medina_OConnor},
which effectively reduces a scalar field theory from $\CP^3_F$ to
$S^4_F$ through a probabilistic penalization method. To this end we
shall briefly review the construction of $S^4_F$.

\subsection{$S^4_F$ revisited}\label{2.1}

We center our attention in the representation theory necessary to
construct the $S^4_F$, further details can be found in
\cite{Medina_OConnor} and \cite{Ramgoolam_0105006}. Consider the
Euclidean gamma matrices of $\mathbb{R}^5$, $\{ \Gamma_a :
~a=1,\ldots,5 \}$, they satisfy the Clifford algebra relations
$\{\Gamma_a ,\Gamma_b \} = 2 \delta_{ab} {\bf 1}$. One may observe
that by defining the operators $\chi_a := \frac{R}{\sqrt{5}}\Gamma_a $
for some real positive number $R$ the relations

\be
\chi_a \chi_a = R^2 {\bf 1}
\ee
are fulfilled. These can be interpreted as the fuzzy analogue of the
embedding equations for $S^{4} \hookrightarrow \mathbb{R}^5$ at the
lowest level of the matrix algebra sequence, that is, the defining
$\Spin(5)$ representation $(\med, \med)$.\footnote{We use everywhere
  the highest-weight vector labeling for representations.} Functions
on $S^4_F$ at this level are given by elements of the form $F = F_0
{\bf 1} + F_a \Gamma_a$ and even at this level they do not form a
closed subalgebra. To solve this difficulty the approach that we
follow, taken in \cite{Medina_OConnor}, is to adopt the full matrix
algebra $Mat_4$. By using the $L$-fold symmetrized tensor product of
the defining representation, $(\frac{L}{2},\frac{L}{2})$, and choosing
as algebra of functions the sequence of matrix algebras formed by the
products $(\frac{L}{2}, \frac{L}{2}) \otimes \overline{(\frac{L}{2},
  \frac{L}{2})}$, the operators
\be \label{defJa}
    J_a  :=    \left( \underbrace{ \Gamma_a {\otimes} \mathbf{1} 
                            {\otimes} \cdots {\otimes} 
                            \mathbf{1}}_{\mbox{\scriptsize L factors}}   +
                            \mathbf{1} {\otimes}  \Gamma_a
                            {\otimes} \cdots {\otimes} 
                            \mathbf{1}   + \cdots +
                            \mathbf{1} {\otimes} \mathbf{1} 
                            {\otimes} \cdots {\otimes} 
                             \Gamma_a
                  \right)_{\rm sym}  
\ee
generalize $\Gamma_a$ to the $L$-th level and satisfy the constraint
\be
J_a J_a = L(L+4){\bf 1}.
\ee
We generalize the matrices $\chi_a$ by defining $X_a :=
\frac{R}{\sqrt{L(L+4)}}J_a$ which satisfy the constraint $X_a X_a =
R^2 {\bf 1}$.  In the large $L$ limit the algebra becomes the
commutative algebra $C^{\infty}(S^4)$ as the commutators $[X_a, X_b]$
vanish in the limit $L \to \infty$ while the constraint remains.
However, at a finite level $L$ the algebra of functions is still not
closed. The procedure presented in \cite{Medina_OConnor} is to enlarge
the algebra of functions to the full matrix algebra and then suppress
the modes which are not associated with the $S^4_F$ degrees of freedom
in the (scalar) fields by giving them a very large excitation
energy. The sequence of matrix algebras obtained is then
$Mat_{d^3_L}$, and we can therefore conceive $S^4_F$ effectively as a
deformed $\CP^3_F$. In what follows we aim to give a geometrical
interpretation of this procedure.

\section{The orbit construction of $\CP^3$}\label{section3} 
In this section we present the construction of $\CP^3$ following
\cite{Medina_OConnor} as $\Spin(6) (\cong \SU(4)$ local isomorphism)
and $\Spin(5)$ orbits and obtain the metric in terms of the
Maurer-Cartan forms of these groups. Hereafter we will specialize to
$N=3$, recalling the Lie algebra isomorphism $spin(6) \cong su(4)$ we
find it convenient to replace the index $\mu = 1,\ldots, 15$ in
(\ref{proj}) by a composite index $\mu = AB$ where each index $A,B =
1,2,\cdots, 6$ and the understanding that they appear only in
antisymmetrized form. In this manner we preserve the use of Einstein's
summation convention. Following \cite{Hamermesh} the algebra
(\ref{algebra}) of the $\Spin(6)$ generators in the fundamental
representation takes the form:\footnote{The relations between gamma
  matrices and the $\Spin(6)$ generators of the ${\bf 4}$
  representation is $\Lambda_{AB} = \med (1 + \Gamma) \frac{1}{4
    \imath} [\Gamma_A, \Gamma_B]$, where $\Gamma = \imath \Gamma_1
  \cdots \Gamma_8 = \Gamma^{\dagger}$ is the chirality and satisfies
  $\Gamma^2 = {\bf 1}$.}
\begin{eqnarray} \label{algebra_lambdas}
    \Lambda_{AB} 
    \Lambda_{CD} &=& A_{AB;CD}\frac{\mathbf{1}}{2}+
                     \frac{1}{4}\epsilon_{ABCDEF} \Lambda_{EF}  \\ \nonumber
                 & & +\frac{\imath}{2}
                     \left( \delta_{AC} \Lambda_{BD}  +
                             \delta_{BD} \Lambda_{AC}  -
                              \delta_{BC} \Lambda_{AD}  -
                                \delta_{AD} \Lambda_{BC}  
                     \right).  
\end{eqnarray}
$ A_{AB;CD}$ is the two-index antisymmetrizer:
\be
  A_{AB;CD}=\med \left( \delta_{AC} \delta_{BD}- \delta_{AD} \delta_{BC}  
                 \right).
\ee
The $d$ and $f$ tensors in (\ref{algebra}) can be read from
(\ref{algebra_lambdas}):
\begin{eqnarray} \nonumber
  d_{ABCDEF} &= & \frac{1}{2} \epsilon_{ABCDEF}, \\ \nonumber
  f_{ABCDEF} &= &  \delta_{AC} A_{BD;EF}- \delta_{AD} A_{BC;EF} +
                             \delta_{BD} A_{AC;EF}- \delta_{BC} A_{AD;EF}.
\end{eqnarray}
The projector $\proj \in Mat_4$ in (\ref{proj}) is expanded
as:\footnote{ We take $\xi_{\mu}\xi_{\mu}=\frac{1}{8}
  n_{AB}n_{AB}$. This is a more convenient normalization for our
  purposes.}
\begin{equation}
 \proj = \frac{1}{4} (\ID + n_{AB} \Lambda_{AB}), \label{proj1}
\end{equation}
the constraints (\ref{constraints}) take the form:

\begin{eqnarray}
  n_{AB} n_{AB}&=&6, \label{restriction1} \\
  \epsilon_{ABCDEF} n_{AB}n_{CD}&=&8n_{EF}. 
                                  \label{restriction2}
\end{eqnarray}
By contractions of (\ref{restriction1})-(\ref{restriction2}) we get
the additional identities:
\begin{eqnarray}
 n_{AC} n_{CB} &=& -\delta_{AB}, \label{restriction3} \\
\epsilon_{ABCDEF} n_{EF}&=& 2(n_{AB} n_{CD} + n_{AD} n_{BC} - n_{AC} n_{BD}), \\
\epsilon_{ABCDEF}n_{AB}n_{CD}n_{EF}&=&48.
\end{eqnarray}
In the coordinate system $\{ n_{AB} \}$ the geometrical objects
(\ref{geometrical}) are
\begin{eqnarray} 
 {\bf P}_{AB;CD} & =& \med A_{AB;CD}+ \frac{1}{8}  \epsilon_{ABCDEF} n_{EF}
                -\frac{1}{4} n_{AB}n_{CD},   \label{proy-P-1} \\
 {\bf J}_{AB;CD} & =& \frac{1}{4}  f_{ABCDEF} n_{EF},  \nonumber \\
&=& \frac{1}{4} (\delta_{AC} n_{BD} - \delta_{AD} n_{BC} + \delta_{BD} n_{AC}
-\delta_{BC} n_{AD}), \nonumber \\
 {\bf K}_{AB;CD} & =& \med \left( {\bf P}_{AB;CD} + \imath {\bf J}_{AB;CD} \right).  \nonumber
\end{eqnarray}
Where, as before, $\bf{J}$, $\bf{P}$ and $\bf{K}$ stand for the
complex structure, metric and K\"ahler structure on $\CP^3$.

A more compact way to express the metric in (\ref{proy-P-1}) is 
\begin{equation}\label{ProyectorP-v2}
{\bf P}_{AB;CD}= \med \left( A_{AB;CD}-Q_{AB;CD} \right)
\end{equation}
where:\footnote{One can then easily check that
    $AQ=QA=Q$,  $Q^2=A$ and $\Tr[Q]=3$.}
\begin{equation}
Q_{AB;CD}=  \med \left(   n_{AC}n_{BD}-  n_{AD}n_{BC} \right).  \label{Qpart-of-P}
\end{equation}
The projector ${\bf P}_{AB;CD}$ has rank $6$, it is the basic projector onto $\mathbb{C}\mathbb{P}^3$. The orthogonal complementary projector in $\mathbb{R}^{15}$ is the rank $9$ projector 
\begin{equation}\label{P_perp}
P^{\perp}_{AB;CD}=\frac{1}{2}(A_{AB,CD}+Q_{AB,CD}) .
\end{equation}

We also have a rank $1$ projector orthogonal to $\CP^3$,
\begin{equation}\label{N_proyector}
N_{AB;CD}=\frac{n_{AB}n_{CD}}{6}.
\end{equation}
%% %%
Notice that ${\bf P}_{AB;CD}$ in (\ref{ProyectorP-v2}) and $N_{AB;CD}$
in (\ref{N_proyector}) give a rank 7 projector. ${\bf P}_{AB;CD} +
N_{AB;CD}$ projects $\mathbb{R}^{15}$ onto $S^7$ which can be viewed
as an $SU(4)$ orbit over $SU(3)$. Since they are orthogonal and
project onto $\CP^3$ and $U(1)$ this $S^7$ admits one squashing
parameter. This is a special case of the more general result that
$S^{2N+1}=SU(N+1)/SU(N)$ and there is always one squashing parameter
associated with the sum of the $\CP^N$ and normal projectors.
% \cite{Dolan_OConnor}.

\subsection{$\CP^3$ as an orbit under $\Spin(6)$  }\label{section4.1}
We give an explicit construction of $\CP^3$ as a $\Spin(6)$ orbit and
analize the induced metric. As the adjoint action of $\Spin(6)$ in the
space of projectors (\ref{cpndef}) is transitive, $\CP^3$ can be
obtained as the $\Spin(6)$ orbit of an appropriate fiducial projector
$\proj^0$: \be \proj=U \proj^0 U^{-1}, \qquad U \in \Spin(6).  \ee For
$\proj^0$ we choose:
\begin{eqnarray}\label{fiducial_1}
  \proj^0&=& \frac{1}{4}(\ID + n_{AB}^0 \Lambda_{AB}), \nonumber \\
     &=& \frac{1}{4}\ID+\frac{1}{2}\left( \Lambda_{12}+ 
                              \Lambda_{34} + \Lambda_{56} \right).
\end{eqnarray}
We call the point corresponding to $\proj^0$ the ``north pole''. 

The projector (\ref{proy-P-1}) plays an essential r\^ole in any differential relations since 
\begin{equation}
dn_{AB}={\bf P}_{AB;CD}dn_{CD}.
\end{equation}
The line element is defined as\footnote{We will typically set $R=1$ for the round $\CP^3$.}
\be\label{so6-line-element}
 ds^2:= \frac{R^2}{8}d n_{AB} d n_{AB} =: \frac{R^2}{8}d n_{AB}^2 =: -\frac{R^2}{8}\Tr (d  \mathcal{N})^2 =\frac{R^2}{8}{\bf P}_{AB;CD}dn_{AB}dn_{CD},\ee
and justifies the appellation metric to the projector ${\bf P}_{AB;CD}$.

The generators $\Lambda_{AB}$ transform as a rank 2 tensor under $\Spin(6)$:
\be \label{xiorbit}
  n_{AB}= R_{AC}R_{BD}n_{CD}^0.
\ee
It may be shown that the line element is given as $d s^2=-\frac{R^2}{4}Tr
\left[R^{-1}d R, \mathcal{N}^0 \right]^2 $,where $\mathcal{N}^0$ is
the matrix with entries $n^0_{AB}$ and we rewrite $R^{-1}d R$ in terms of left invariant Maurer-Cartan forms of $\Spin(6)$ \footnote{It would be more natural to use right-invariant 
Maurer-Cartan forms (see Appendix B), since these are dual to the vector fields ${\cal L}_{AB}$ discussed below, but both will be equivalent at the north pole, the resulting right-invariant expressions 
are equivalent to replacing the projectors at the north pole by those at a generic point.}: $R^{-1}d R=:-\imath e_{AB}T_{AB}$, where $T_{AB}$ are the generators of the vector
representation.\footnote{The normalization of the generators $T_{AB}$ is such that they satisfy the same Lie algebra as $\Lambda_{AB}$ with identical structure constants, their matrix 
elements being $(T_{AB})_{IJ} = -\imath (\delta_{AI}\delta_{BJ} -\delta_{AJ}\delta_{BI})$.} The line element is hence:

\begin{eqnarray}\label{line_element_spin6}
 ds^2 &=& 4R^2e_{AB}{\bf P}^0_{AB;CD}e_{CD} = 4 R^2 \Big( (e_{13}-e_{24})^2 + (e_{14}+e_{23})^2 + (e_{15}-e_{26})^2 +  \nonumber \\
     & & (e_{16}+e_{25})^2 + (e_{35}-e_{46})^2 + (e_{36}+e_{45})^2 \Big). 
\end{eqnarray}
It becomes apparent from (\ref{line_element_spin6}) that the orbit is
a six dimensional space, as expected for $\CP^3$. It is possible to
obtain the isotropy subgroup by looking at the combinations of forms
$e_{AB}$ which do not appear in the metric; the corresponding
combinations of generators span the isotropy subalgebra. In the
$SU(4)$ formulation the isotropy group is easily identified as
$S(U(3)\times U(1))$. We obtain a coset space realization for $\CP^3$
\be \CP^3=SU(4)/S(U(3) \times U(1)). \ee

\subsection{$\CP^3$ as an orbit under $\Spin(5)$}\label{3.2}
Observe that $\{ \Lambda_{ab},~~a,b=1,...,5 \}$ generate the $spin(5)$ subalgebra of  $spin(6)$ while $\Lambda_{a6}$ transforms as a vector under $\Spin(5)$. 
We define $\Lambda_a := \Lambda_{a6}$ so we can write the projector (\ref{proj}) as:
\be
  \proj=\frac{1}{4}\ID+ \frac{n_a}{2} \Lambda_a +\frac{n_{ab}}{4} \Lambda_{ab}, 
\ee
the projector (\ref{fiducial_1}) takes the form:
\be \label{projspin5}
  \proj^0=\frac{1}{4}\ID+ \frac{1}{2} \Lambda_5 +\frac{1}{2}\left(
                    \Lambda_{12} + \Lambda_{34}\right).
\ee
The action of $\Spin(5)$ on the space of projectors (\ref{cpndef}) is also transitive, hence we obtain $\CP^3$ as an orbit of (\ref{fiducial_1}) or (\ref{projspin5}) under $\Spin(5)$.

The function algebra of $\CP^3$ is now built from the two $Spin(5)$ representations $n_a$ which carries the 5-dimensional representation and $n_{ab}$ which carries the 10-dimensional representation.
The $SO(6)$ invariant line element (\ref{so6-line-element}) can therefore 
be deformed to:
\be\label{dbars2}
   d\bar{s}^2= \alpha dn_a^2 + \beta dn_{ab}^2 .
\ee
We will leave $\alpha,\beta$ undetermined for the moment, so that this
is the most general induced $\Spin(5)$-invariant line element, we will
come back to this point at the end of this section and in section
\ref{section5}.  It is now possible to write
\begin{eqnarray}
d n_{ab}^2 &=&-Tr \left[R^{-1}d R, n^0  \right]^2, \\
dn_a^2&=&  |R^{-1}dR n^0_v|^2,
\end{eqnarray}
where $n^0$ stands for the matrix of coefficients $n_{ab}^0$, and $n^0_v$ for the column vector with components $n_a^0$.

As before we express the line element in terms of the $\Spin(5)$ Maurer-Cartan forms $R^{-1}d R:= -\imath e_{ab}T_{ab}$ where $T_{ab}$ are the generators of the $spin(5)$ 
subalgebra in the vector representation.

We obtain for the line element:

\begin{eqnarray} \label{line-element-spin5}
  d\bar{s}^2 &=& \Big( \frac{R^2_{S^4}}{2}{\mathbb P}^0_{ab,cd} + R^2_{S^2} {\mathbb X}^0_{ab,cd}  \Big) e_{ab} e_{cd}  \\ \nonumber
  &=&  ( \alpha + 2 \beta) \Big( e_{15}^2 + e_{25}^2 + e_{35}^2 + e_{45}^2 \Big) + 4\beta [ (e_{14} + e_{23} )^2 + (e_{13} - e_{24})^2] 
\end{eqnarray} 

From (\ref{line-element-spin5}) we can observe two interesting features: First, the isotropy subgroup can be constructed as before giving the following 
coset space realization $\CP^3 = \Spin(5)/[U(1)\times SU(2)]$ and second, the space is locally of the form $S^2 \times S^4$; $\CP^3$ is indeed a fibre bundle with base space $S^4$ and 
fibre $S^2$. The constants $\alpha, \beta$ can now be reinterpeted in terms of the squared radii of these spheres: $R_{S^4}^2 =  \alpha + 2\beta$ and $R_{S^2}^2 = 4 \beta$ and the line 
element (\ref{dbars2}) can be written in the form
\begin{equation}
d\bar{s}^2=R_{S^4}^2dn_a^2+\frac{R_{S^2}^2}{4}(dn_{ab}^2-2 dn_a^2) .
\end{equation}
Furthermore using 
\begin{equation}
dn_a=n_{ac}n_bdn_{cb} 
\end{equation}
one can extract the projector $\mathbb{X}_{ab;cd}$ onto
the $S^2$ fibre. This and related projectors are discussed in 
section \ref{section5.1} below. The line element can therefore be written as 
\begin{equation}
d\bar{s}^2=R_{S^4}^2 dn_a^2+\frac{R_{S^2}^2}{4}{\mathbb X}_{ab;cd} dn_{ab}dn_{cd} .
\end{equation}
If we restrict the Maurer-Cartan forms in (\ref{line_element_spin6})
to $SO(5)$ we see that we recover the line element (\ref{line-element-spin5})
with $R_{S^4}^2 = R_{S^2}^2=R^2$.

\section{Scalar field theory on $S^4_F$ revisited}\label{section4}
As it was stated in section 3, $\CP^3$ can be obtained as a $\Spin(6)$ or $\Spin(5)$ orbit. In order to specify the geometry all that is needed is to define a Laplacian. In principle we 
can choose the $\Spin(6)$ or $\Spin(5)$ quadratic Casimir operators, or even a more general choice: an interpolation between both of them:
\[
  k_1 \CAS^{\Spin(6)}   + k_2 \CAS^{\Spin(5)}.
\]

In \cite{Medina_OConnor} a prescription for a generic scalar field theory on fuzzy $\CP^3$ was given, the expression for the action reads
\begin{equation}  
S[\Phi]=\frac{\Tr}{d_L}\left(\frac{1}{2}\Phi\Delta_h \Phi +  V[\Phi]\right),
\end{equation}
where the full Laplacian is

\be
 \Delta_h   =  \frac{1}{8 R^2}\left( \CAS^{\Spin(6)}  
+ h ( 2\CAS^{\Spin(5)}-\CAS^{\Spin(6)}) \right).
\label{Delta_cas}
\ee
As mentioned in section 2, the algebra of functions on $\CP^3$ is approximated by a sequence of matrix algebras of dimension $d^3_L= \frac{(L+1)(L+2)(L+3)}{6}$.

The quadratic Casimir operators can be written using the adjoint action of the corresponding generators
\begin{eqnarray}
 \med \CAS^{\Spin(6)} &=& \left(ad \gen_{AB}\right)^2 ,
\label{so6_cas} \label{J_ABgenerators} \\
 \med \CAS^{\Spin(5)} &=& \left(ad \gen_{ab}\right)^2 . \label{J_ab_generators}
\label{so5_cas}
\end{eqnarray}
The normalization of  $\mathcal{J}$ in (\ref{J_ABgenerators})  has been chosen so that in the fundamental representation $\mathcal{J}_{AB} = \med \Lambda_{AB}$.  In the same manner we 
have in (\ref{J_ab_generators}) $\mathcal{J}_{ab} = \med \Lambda_{ab}$ for the fundamental representation.
For the $\Spin(6)$ generators we use those in the $L$-fold symmetric tensor product representation $(\frac{L}{2},\frac{L}{2},\frac{L}{2})$
with the same dimension $d^3_L$, for $\Spin(5)$ we use the generators of the $(\frac{L}{2},\frac{L}{2} )$ representation, whose dimension is also $d^3_L$.

The choice (\ref{Delta_cas}) for the Laplacian can be understood analyzing the effect of the term 
\begin{equation}
   \Delta_I = \frac{1}{8R^2}\left(2\CAS^{\Spin(5)}-\CAS^{\Spin(6)} \right) \label{Delta_h_Cas}
\end{equation}
on $S_F^4$ modes.
After an analysis of the representation content for (\ref{matalgebra}) it was proved in \cite{Medina_OConnor} that $\Delta_I$ is a strictly positive operator for the non-$S^4_F$ modes 
and has as its kernel precisely the $S^4_F$ modes.

The mechanism is one of probabilistic penalization as the probability of a field configuration $\Phi$ can be separated into
 
\begin{equation}
 \Proba[\Phi]=\frac{{\rm e}^{-S[\Phi]-hS_I[\Phi]}}{Z}
\label{prob_of_config}
\end{equation}
where 
\begin{equation}
Z=\int d[\Phi] {\rm e}^{-S[\Phi]-hS_I[\Phi]}
\label{partition_fn}
\end{equation}
is the partition function of the model. Taking the limit $h \to \infty$ makes the non-$S^4_F$ modes unreachable. The final result is that the $\CP^3_F$ field configurations not 
related to $S^4_F$ are dynamically supressed in this limit.

\section{Geometric analysis of the supression mechanism}\label{section5}
The product of matrices together with a map to functions induces a noncommutative product on functions, this is the {\em *-product}  \cite{starprod_CPN}. This useful tool allows us to 
access the commutative limit explicitely.
Let $\widehat{M_1}$, $\widehat{M_2}$ be two matrices of dimension $d^3_L$ and  $M_1(n)$, $M_2(n)$ be the corresponding functions obtained by the mapping:
\be
  M_1(n):=Tr \left( {\proj}_L(n) \widehat{M_1} \right)
\ee
$\proj_L(n) $ is constructed by taking the $L$-fold tensor product of $\proj$ defined in (\ref{proj}), it provides a map to functions at the level $L$. 
%Notice $ \proj_L(\xi) $ carries the coordinates $\xi$. 

The *-product is then defined through:
\be
  \left( M_1 * M_2 \right) (n):=Tr \left( \proj_L(n) 
        \widehat{M_1} \widehat{M_2} \right).
\ee
For $\CP^{N}$ the *-product can be written as a finite series of derivatives on the coordinates $n_{AB}$, for our purposes we will use the prescription given in \cite{starprod_CPN}.
\begin{equation}
  n_{AB}:= 4 Tr\left( \proj_L(n) \gen_{AB} \right).
\end{equation}
The commutator of $\gen_{AB}$ maps into the right-invariant vector fields:

\begin{eqnarray}
 \calL_{AB} M(n)&:=&Tr \left(\proj_L (n) \left[\gen_{AB},\widehat{M}\right] \right) \\
                  &= &2\imath {\bf J}_{AB;CD}\partial_{CD}M(n)
\end{eqnarray} 

%The quadratic Casimir operators are defined as:
%\begin{eqnarray}
% [\gen_{AB},[\gen_{AB},\cdot ]]&=&C_2^{SO(6)},  \label{cc6}  \\   
% \left[\gen_{ab},[\gen_{ab}, \cdot ]\right]&=&C_2^{SO(5)} \cdot  \label{cc5}
%\end{eqnarray}
The images of (\ref{so6_cas})-(\ref{so5_cas}) under the *-product map are:

\begin{eqnarray}
  \med \CAS^{\Spin(6)} \widehat{M} =\left[\gen_{AB},[\gen_{AB},\widehat{M}]\right] \longrightarrow 
             \mathcal{C}^{(6)} M(n)=-4\kappa_6 M(n) ,    \\ 
  \med \CAS^{\Spin(5)} \widehat{M}=\left[\gen_{ab},[\gen_{ab},\widehat{M}]\right]
  \longrightarrow  \mathcal{C}^{(5)} M(n)=-4 \kappa_5 M(n) .
\end{eqnarray}
where
\begin{eqnarray}
  \kappa_6 &=&  {\bf J}_{AB,CD} \partial_{CD}\left( {\bf J}_{AB,EF} 
                               \partial_{EF} \right)           \\                          &=& {\bf P}_{CD;EF}\partial_{CD}\partial_{EF}+
                {\bf J}_{AB;CD}(\partial_{CD} {\bf J}_{AB;EF})\partial_{EF}  \\             \kappa_5 &=&  {\bf J}_{ab,CD} \partial_{CD}\left( {\bf J}_{ab,EF} 
                               \partial_{EF} \right)    
\end{eqnarray}

%A very useful expression to simplify the calculations is:
%\begin{equation}
%  J_{AB;CD}\partial_{CD}=\sqrt{2}\left( \xi_{AC}\partial_{CB}- \xi_{BC}\partial_{CA} \right)
%\end{equation}

Now, we are interested in extracting the metric tensor comparing the relevant continuous Laplacian with the general form:
\begin{eqnarray}
 -\calL^2&=& \frac{1}{\sqrt{G}}
              \partial_{\mu}\left(\sqrt{G}G^{\mu \nu} \partial_{\nu} \right) \\
        &=& G^{\mu \nu}  \partial_{\mu} \partial_{\nu} +
              \left(\partial_{\mu}G^{\mu \nu} \right)\partial_{\nu} +
              \frac{1}{\sqrt{G}}  G^{\mu \nu}
              \left( \partial_{\mu} \sqrt{G}  \right)\partial_{\nu}.
\end{eqnarray}
When we retain the full $\Spin(6)$-symmetry we have the Laplacian $\mathcal{C}^{(6)}$, the associated metric tensor is just ${\bf P}_{AB;CD}$ as can be seen from a  straightforward 
calculation of $\kappa_6$:

\begin{equation}
   \kappa_6=\frac{1}{2 }\partial^2_{AB}+
            \med n_{CB}n_{DA}\partial_{AB}\partial_{CD}-
            n_{AB}\partial_{AB}.
\end{equation}
If one retains only the $\Spin(5)$ symmetry, the following expressions are found

\begin{eqnarray}
\kappa_6 &=& \med \partial_{ab}^2 + \partial_a^2 + \med n_{cb}n_{da}\partial_{ab}\partial_{cd} + 2 n_a n_{bc} \partial_{ab} \partial_c - n_a n_b \partial_a \partial_b \\ \nonumber 
& & - n_{ab}\partial_{ab} - 2 n_a \partial_a, \\
\kappa_5 &=& \med \partial_{ab}^2 + \med \partial_a^2 - \med n_a n_b \partial_{ac}\partial_{bc} + \med n_{ca}n_{bd}\partial_{ab}\partial_{cd} - n_a n_{cb}\partial_{ab}\partial_c \\ \nonumber & & - \med n_a n_b \partial_a \partial_b - \frac{3}{4} n_{ab} \partial_{ab} - n_a \partial_a.
\end{eqnarray}

\subsection{The vertical and horizontal projectors}\label{section5.1}

The vertical and horizontal projectors can be constructed explicitely in our coordinate system in a $\Spin(5)$-covariant manner. The coordinates $n_{AB}$ break up under $\Spin(5)$ as $n_{ab}$ and $n_a$. 
These are the basic objects we will need to build projectors. From (\ref{restriction3})  they satisfy
\begin{eqnarray}
n_{ac}n_{bc}&=& \delta_{ab}-n_an_b =P_{ab}, \\
 n_an_{ab}&=&0, \\
n_an_a&=&1,
\end{eqnarray}
%% %%
$P_{ab}$ is a rank 4 projector, it projects $\mathbb{R}^5 \mapsto S^4$, the usual continuum embeding
of $S^4$ in $\mathbb{R}^5$, and its orthogonal complement is $ P^{\perp}_{ab}=\delta_{ab} - P_{ab}=n_a n_b$. Defining

\be
Q_{ab,c}=\frac{1}{2}(n_{ac}n_b-n_an_{bc}), \quad Q_{ab;cd}=\frac{1}{2}(n_{ac}n_{bd}-n_{ad}n_{bc}) 
\ee
we observe that 
\be 
2 Q_{ab,e}Q_{cd,e}= {\mathbb{P}}_{ab;cd}  = \frac{1}{2}(\delta_{ac}P^{\perp}_{bd} -\delta_{ad}P^{\perp}_{cb}+\delta_{bd}P^{\perp}_{ac}-\delta_{bc}P^{\perp}_{ad}).
\ee
Where $\mathbb{P}_{ab,cd}$ projects $\mathbb{R}^{10} \mapsto S^4$, it is therefore the metric on $S^4$, the horizontal projector. We can then define the projectors $\mathbb{X}$ and $Y$: 

\begin{eqnarray}
\mathbb{X}_{ab;cd} &=& \frac{1}{2}(A_{ab;cd}-{\mathbb{P }}_{ab;cd}-Q_{ab;cd}), \label{PROY-X} \\
Y_{ab;cd} &=& \frac{1}{2}(A_{ab;cd}-{\mathbb{P }}_{ab;cd}+Q_{ab;cd}). \label{PROY-Y}
\end{eqnarray}
Notice the ranks $\Tr[\mathbb{X}]=2$ and $\Tr[Y]=4$.

To see that these are orthogonal projectors one needs to observe that
\be
Q_{ab;cd}Q_{cd;ef}=A_{ab;ef}-{\mathbb{P }}_{ab;ef}.
\ee
The tensor $\mathbb{X}_{ab;cd}$ is the projector onto the fibres of $\mathbb{C}\mathbb{P}^3$ as an $S^2$ bundle over $S^4$, it is the vertical projector. $\mathbb{X},Y$ and $\mathbb{P}$ 
are complementary and add up to the identity in $\mathbb{R}^{10}$, $A_{ab,cd}$. It is straightforward to write the projector to the bundle, 
${\bf P}_{ab,cd} : \mathbb{R}^{10} \mapsto \CP^3$ as

\be
{\bf P}_{ab,cd} := \mathbb{X}_{ab;cd} + {\mathbb{P}}_{ab;cd}  =\med (A_{ab;cd} + {\mathbb{P}}_{ab;cd} - Q_{ab;cd}).
\ee
Using these projectors we construct an ansatz for the metric of the squashed $\CP^3$.

For completeness we also give the complex structure of $\CP^3$ in the $\Spin(5)$ formulation, we start by defining 

\begin{eqnarray} \nonumber
T_{ab;cd} &=& \frac{1}{4}(P_{ac} n_{bd} - P_{ad} n_{bc} + P_{bd} n_{ac} - P_{bc} n_{ad} ) \\ \nonumber
\tilde{T}_{ab;cd} &=& \frac{1}{2}(P_{ac}^\perp n_{bd} - P_{ad}^\perp n_{bc} + P_{bd}^\perp n_{ac} - P_{bc}^\perp n_{ad} )
\end{eqnarray}
and noting that 
\begin{eqnarray} \nonumber
T_{ab;ef}T_{ef;cd}& =&  - \mathbb{X}_{ab;cd}\ ,\qquad\qquad   T_{ab;ef}T_{ef;gh}T_{gh;cd}=-T_{ab;cd}\ , \\  
 \tilde{T}_{ab;ef}\tilde{T}_{ef;cd} &=& - \mathbb{P}_{ab;cd} \qquad\ \hbox{and}\quad  \tilde{T}_{ab;ef}\tilde{T}_{ef;gh}\tilde{T}_{gh;cd}=-\tilde{T}_{ab;cd}\ .  \nonumber
\end{eqnarray}
One constructs ${\bf J} = T + \tilde{T}$ resulting into:\footnote{It should be mentioned that since $S^4$ does not even admit an almost-complex structure, so 
$\tilde{T}$ is not a complex structure on it.}

\be
{\bf J}_{ab;cd} = \frac{1}{4}(\delta_{ac} n_{bd} - \delta_{ad} n_{bc} +\delta_{bd} n_{ac} - \delta_{bc} n_{ad} + P_{ac}^\perp n_{bd}-P_{ad}^\perp n_{bc} + P_{bd}^\perp n_{ac} 
- P_{bc}^\perp n_{ad} ).
\ee
It is easy to prove that 

\be
{\bf J}^2 = - {\bf P}. 
\ee
We return now to the discussion regarding the Laplacian. For the deformed case, which possesses only $\Spin(5)$ symmetry, the 
corresponding Laplacian acting on functions is $\calL^2_h$, we have

\begin{equation}
\Delta_h= \frac{1}{8 R^2} \left( \CAS^{\Spin(6)} +h\left(2\CAS^{\Spin(5)}-\CAS^{\Spin(6)}  \right) \right)
\end{equation}

and the mapping is:
\begin{equation}
  Tr\left(  \proj_L (n) \Delta_h  \widehat{M} \right)=:\frac{1}{R^2} \calL^2_h M(n),
\end{equation}
then
\begin{equation}
  -\calL^2_h= \kappa_6 +h\left(2\kappa_5-\kappa_6  \right).
\end{equation}
Our ansatz for the metric tensor related to $ \calL^2_h$ is the following:\footnote{ Double indices in ${\bf P}_{ab;cd}$ and $\mathbb{X}_{ab;cd}$ are raised and lowered using 
$A^{ab;cd}$ and $A_{ab;cd}$, the complex structure ${\bf J}$ is thus an up-down tensor.}

\begin{equation}
 G^{ab;cd}= \frac{2\mathbb{P}^{ab;cd} + (h+1)\mathbb{X}^{ab;cd}}{R^2} = \frac{2\mathbb{P}^{ab;cd}}{R^2} + \frac{\mathbb{X}^{ab;cd}}{R^2_{S^2}} ; 
\label{metricG}
\end{equation}
the tensor $\mathbb{X}_{ab;cd} = \mathbb{X}_{cd;ab}$ is recovered 
from the combination 

\begin{equation}
 2\kappa_5-\kappa_6=\frac{1}{2}\partial_{ab}^2-\med n_{ab}n_{cd}\partial_{ac}\partial_{bd}
                   - n_{a}n_{b}\partial_{ac}\partial_{bc}-  \med n_{ab}\partial_{ab}
 \label{exterm}
\end{equation}
by comparing the term in second derivatives in (\ref{exterm}) against the $h$-dependent term in (\ref{metricG}), $ \mathbb{X}_{ab;cd}\partial_{ab}\partial_{cd}$, 
and we find that $\mathbb{X}$ thus obtained is indeed the fibre metric we had previously identified in (\ref{PROY-X}), i.e.

\begin{eqnarray}
   \mathbb{X}_{ab;cd}&=& \frac{1}{2} A_{ab;cd}-  \frac{1}{4} \left(\delta_{ac} n_{b}n_{d} -
                        \delta_{ad}   n_{b}n_{c}+ 
                        \delta_{bd}   n_{a}n_{c}-  
                        \delta_{bc}   n_{a}n_{d} \right)\nonumber \\  & &
                - \frac{1}{4} \left(n_{ac}n_{bd}-
                                             n_{ad} n_{bc}  
                                       \right).  \label{X-projector}
\end{eqnarray}
In order to invert the metric tensor (\ref{metricG}) we observe that ${\bf P}\mathbb{X} = \mathbb{X}{\bf P} = \mathbb{X}$, hence the covariant metric tensor is a linear combination of 
$\mathbb{X}$ and ${\bf P}$, in fact
\begin{equation} \label{covariantmetric}
   G_{ab;cd}= R^2( \frac{\mathbb{P}_{ab,cd}}{2} + \frac{1}{h+1}\mathbb{X}_{ab;cd}) = \frac{R^2 \mathbb{P}_{ab,cd}}{2} + R^2_{S^2} \mathbb{X}_{ab;cd}
\end{equation}
satisfies the required condition: $G^{ab;cd}G_{cd;ef}= {\bf P}^{ab}_{ef}$.

\section{Fermion fields}\label{section6}
A fuzzy four-dimensional fermion field has the representation content:
\be
 \Psi \in (\med ,\med, \med) \otimes (\frac{L}{2},\frac{L}{2},\frac{L}{2}) \otimes (\frac{L}{2},\frac{L}{2}, -\frac{L}{2}).
\ee
It is shown in \cite{Medina_OConnor} that the algebra of fuzzy functions decomposes as

\be
 (\frac{L}{2},\frac{L}{2},\frac{L}{2}) \otimes (\frac{L}{2},\frac{L}{2}, -\frac{L}{2})   = \bigoplus_{n=0}^{L} (n,n,0),
\ee
hence, the relevant decomposition is

\begin{eqnarray} \nonumber \label{spinor}
(\med, \med , \med) \otimes (n,n,0) &=& \underbrace{(n+ \med , n+ \med , \med)}_{n\geq 0} \oplus \underbrace{(n+\med, n-\med, -\med)\oplus (n-\med,n-\med,\med)}_{n\geq 1}\\ 
&: =& D^n_+ \oplus D^n_0 \oplus D^n_- .
\end{eqnarray}
The restrictions below show when these representations appear in the decomposition. The spinor field decomposes into components $\Psi = \Psi_+ \oplus \Psi_0 \oplus \Psi_- $.

For a Dirac operator on $S^4_F$  we propose the linear spinor 
operator in the spirit of \cite{Dolan_OConnor} given by the ansatz
\be \label{Dirac}
\D_{\tilde h} = \sigma_{AB} [\gen_{AB}, \cdot ] + 2 + {\tilde h} \left( 2\sigma_{ab} [\gen_{ab}, \cdot ]-\sigma_{AB} [\gen_{AB}, \cdot ] \right),
\ee
where $\sigma_{AB}$ are the $\Spin(6)$ generators in the fundamental
representation $(\med, \med, \med)$ and $\sigma_{ab}$ are the
corresponding $\Spin(5)$ generators. The operator $\D_{\tilde h}$ can be
expressed in terms of the differences of Casimir operators
$\mathfrak{C}_2 :=2 ([\gen, \cdot] + \frac{\sigma}{2})^2$ and 
$C_2=2[\gen,\cdot]^2$ . 

By ``completing the square'' we may rewrite the operator $\D_{\tilde h}$ purely in terms of quadratic Casimir operators. 
Note that $[\mathfrak{C}_2^{\Spin(5)},\mathfrak{C}_2^{\Spin(6)}] = 0$ as can be readily verified by expanding out $\mathfrak{C}_2^{\Spin(6)}$ in $\Spin(5)$ 
indices in a $\Spin(5)$ invariant manner. It is then clear that both Casimir operators can be simultaneously diagonalized in the appropriate basis. In order to compute the spectrum of 
the given operator (\ref{Dirac}) we use the following reductions under $\Spin(5)$

\begin{eqnarray}
(n+\med, n-\med, -\med) &=& \bigoplus_{m=0}^{n-1} \left( (n+\med, m +\med )\oplus(n-\med, m+\med)\right),   \\
(n+\med, n+ \med, \med) &=& \bigoplus_{m=0}^{n}(n+\med,m +\med).
\end{eqnarray}
Wherefrom we find the decompositions

\begin{eqnarray}
\Psi_+ &=&  \bigoplus_{m=0}^{n} \Psi^{ (n + \med, n+ \med ,\med)}_{(n+\med, m+\med),+},   \label{spinor-spin5-1} \\
\Psi_0 &=&   \bigoplus_{m=0}^{n-1} \left( \Psi^{(n +\med , n-\med, -\med)}_{(n +\med ,m + \med),0} \oplus \Psi^{(n +\med , n-\med, -\med)}_{(n-\med,m + \med),0} \right), 
\label{spinor-spin5-2}  \\
\Psi_- &=&  \bigoplus_{m=0}^{n-1} \Psi^{(n-\med, n-\med, \med)}_{(n-\med, m+\med),-}~~. \label{spinor-spin5-3} 
\end{eqnarray}

The spectrum of the operator $\D_h$ corresponding to the component
$\Psi_0$ has no counterpart in the known spectrum for the Dirac
operator on $S^4$, therefore this component corresponds to degrees of
freedom extraneous to the $S^4$ and it will, in fact,  be completely
supressed by our dynamical mechanism.  The contributions to $\Psi_+$ and $\Psi_-$ in the kernel of $\D_I = 2\sigma_{ab} [\gen_{ab}, \cdot ]-\sigma_{AB} [\gen_{AB}, \cdot ]$ reproduce a cutoff version of the canonical spectrum of Dirac operator on 
the round $S^4$.

In detail we have the following eigenvalues, 
calculated with the expressions found in
appendix \ref{A}

\begin{eqnarray} \nonumber
\D_{\tilde h} \Psi^{ (n + \med, n+ \med ,\med)}_{(n+\med, m+\med),+}  &=& ( n + 2 + {\tilde h} m )  \Psi^{ (n + \med, n+ \med ,\med)}_{(n+\med, m+\med),+}   \qquad n \geq 0, \\
%\nonumber
\D_{\tilde h} \Psi^{(n-\med, n-\med, \med)}_{(n-\med, m+\med),-} &=&  (-n-1 +  {\tilde h} m) \Psi^{(n-\med, n-\med, \med)}_{(n-\med, m+\med),-}  ~\quad n \geq 1 ,\\ 
\nonumber
\D_h \Psi^{(n+\med,n-\med,-\med)}_{(n+\med,m+\med),0} &=& (1+{\tilde h}(n+m+1))\Psi^{(n+\med,n-\med,-\med)}_{(n+\med,m+\med),0} \qquad n \geq 1, \\ 
\nonumber
\D_h \Psi^{(n+\med,n-\med,-\med)}_{(n-\med,m+\med),0} &=& (1+{\tilde h}(m-n+2))\Psi^{(n+\med,n-\med,-\med)}_{(n-\med,m+\med),0} \qquad n \geq 1.
\end{eqnarray} 
In the large ${\tilde h}$ limit the portion of the spectrum not in the kernel of $\D_I$ 
is sent to infinity, the remaining low lying spectrum coincides with the spectrum of the Dirac operator on $S^4$ 
up to a truncation \cite{Balachandran:2003}, namely 
\be \nonumber
\{ \pm (n+2) ~:~ n = 0,1,\cdots, L-1 \}\cup \{ L + 2 \}, ~ deg(n+2) = \frac{2 (n+1)(n+2)(n+3)}{3}. 
\ee
The degeneracies have been calculated using the formulae in appendix \ref{A}, clearly one has $deg(n+2) = \dim (n+\med, \med)$.

A fermionic action may be now be written for a free spinor field with
mass $M$ as \be S_{\Psi} = \frac{\Tr}{d^3_N} \left( \bar{\Psi} (\D_{\tilde{h}} +
M )\Psi \right).  \ee We remark that the deformed spinor operator
$\D_{\tilde{h}}$ is not a Dirac operator on $\CP^3$ with a squashed metric, our
purpose here is to find a suitable operator for Fermions on fuzzy
$S^4$.  The operator we have found has similarities to higher spin
Dirac operators introduced in \cite{Padmanabhan}.  As in the case of
the scalar theory the statistical penalization mechanism will suppress
the functional degrees of freedom in the spinor field $\Psi$ which are
not associated to $S^4_F$.

One can check that when maped to functions the operator $D_I$ is mapped to 
$\sigma_{ab}\mathbb{X}_{ab;cd}\partial_{cd}$ and since $n_e$ is in the kernel of this 
operator any function of $n_e$ is in the kernel. It sees only the dependence on $n_{ab}$. The parameter ${\tilde h}$ is similarly related to the radius of the $S^2$ fibres and for large ${\tilde h}$ we are shrinking the fibres relative to the $S^4$ base. 

\section{Conclusions}\label{section7}

We review the construction of fuzzy $\CP^3$ presented in
\cite{starprod_CPN}. The main motivation to discretize this $6$
dimentional space is due to its relation to $S^4$, a compactification
of $\mathbb{R}^4$.

The standard construction of $~\CP^3$ involves $\Spin(6)$ symmetry,
giving as result a ``round'' version of $\CP^3$. We gave a different
construction of $\CP^3$ and its fuzzy version as a $\Spin(5)$ orbit
where the local structure $S^2 \times S^4$ is manifiest. The isotropy
group was found to be $SU(2)\times U(1)$. Following the results
obtained in \cite{Medina_OConnor} in which a convenient interpolation
of the $\Spin(6)$ and $\Spin(5)$ quadratic Casimirs was introduced as
the Laplacian, we interpret the deformation parameter $h$ introduced
in \cite{Medina_OConnor} in terms of the radii of a squashed $\CP^3$.
From the point of view of a scalar field theory this procedure can be
interpreted as a Kaluza-Klein construction, where the entire space is
non-trivial fibre bundle with base $S^4$ and fibre $S^2$, and in the
large $h$ limit the radius of the $S^2$ fibres is sent to zero.

Along the way we constructed the complex structure of $\CP^3$ as 
a $Spin(5)$ orbit. The square of the complex sturcture gives minus the 
$\CP^3$ projector and it naturally splits into parts which give the 
$S^4$ base and $S^2$ fibres.

Using *-product map techniques we have presented an explicit manner to
extract the metric of the space under consideration from its
Laplacian. The explicit form of the deformed metric tensor
$G_{\mu\nu}$ was obtained. Examining the resulting line element $d
s^2$ we found the ratio between radii:

\[
\frac{R_{S^2}}{R} = \frac{1}{\sqrt{(1+h)}} .
\]
The limit $h\rightarrow\infty$ corresponds to shrinking the $S^2$
fibres down to zero size, while the limit $h\rightarrow -1$ makes the
fibres infinitely large.
 
We have also proposed a linear spinorial operator on $S^4_F$, based on
the same geometric structure as the scalar case, and identified the
relevant spinor subspaces that contain the correct spectrum of the
Dirac operator on $S^4$, up to a truncation. This operator acts on
four component spinors and does not correspond to a Dirac operator on
$\CP^3$, though it is a well defined first order operator on $\CP^3$
and its fuzzy version. As with scalar fields, spinor fields on $S^4_F$
have additional degrees of freedom in the construction, however all
become of arbitrarily large mass as the parameter $\tilde{h}$ is sent to
infinity and so are dynamically suppressed.

\vspace{1cm}

\noindent {\large \bf Acknowledgements }

\noindent It is a pleasure to thank A.P. Balachandran, Xavier Martin and Peter Pre\v{s}najder  for helpful discussions. J. Medina is grateful to DIAS where part of this work was carried 
out, the support of COFFA (IPN, Mexico) is acknowledged. I. Huet thanks the TPI for hospitality and support, his work was funded through the DFG grants Gi 328/3-2 and SFB-TR18.

\appendix
\section{Casimir operators and dimensions } \label{A}

The quadratic Casimir operators for $\Spin(6)$ and $\Spin(5)$ found in \cite{Popov, Perelomov1, Fulton} were used,

\begin{eqnarray}
\CAS^{\Spin(6)} (m_1 , m_2, m_3) &=& m_1^2 + m_2^2 + m_3^2 + 4m_1 + 2m_2 , \\ 
\CAS^{\Spin(5)} (m_1,m_2) & = & m_1(m_1 + 3) + m_2 (m_2 +1), 
\end{eqnarray}
which for the involved representations amount to
\begin{eqnarray}
\CAS^{\Spin(6)} (n,n,0) & = & 2n (n+3) ,\\
\CAS^{\Spin(6)} (n + \med, n+ \med, \med) &=& 2n(n+4) + \frac{15}{4},  \\
\CAS^{\Spin(6)} (n + \med , n-\med , -\med) & = & 2n(n+3) + \frac{7}{4},\\
\CAS^{\Spin(5)} (n + \med, m + \med) &=& n(n+4) + m(m+2) + \frac{5}{2},\\
\CAS^{\Spin(5)} (n -\med , m + \med) &=& n(n+2) + m(m+2) - \med.
\end{eqnarray}
We rewrite the square taking into account the following normalization:

\begin{eqnarray} 
(\frac{\sigma_{AB}}{2})^2  &= & \med \CAS^{\Spin(6)} (\med, \med,\med) =\frac{15}{8},\\ 
\lbrack \mathcal{J}_{AB}, \lbrack \mathcal{J}_{AB}, \cdot \rbrack \rbrack  &=& \med \bigoplus_{n=0}^L \CAS^{\Spin(6)} (n,n,0).
\end{eqnarray}
Some useful formulae for the dimensions of representations we deal with are

\begin{eqnarray}
\dim (m_1, m_2, m_3) &=& \frac{1}{12} (m_1^2 - m_2^2 +4m_1 -2m_2 +3)   \\ \nonumber
               & &   \times (m_1^2 - m_3^2 + 4m_1 + 4)(m_2^2 - m_3^2 + 2m_2 +1) \\
\dim (m_1, m_2) &=& \frac{1}{6}(m_1^2 -m_2^2 + 3m_1 - m_2 + 2  )\\ \nonumber
                & & \times (2m_1 + 3)(2m_2 +1).
\end{eqnarray}
\begin{eqnarray}
\dim (n,0) &=& \frac{1}{6}(n+1)(n+2)(2n+3)  ,\\ 
\dim (n,n,0)&=& \frac{1}{12} (n+1)^2 (n+2)^2(2n+3) , \\
\dim (n+\med, n+\med, \med) &=& \frac{1}{6} (n+1)(n+2)^3 (n+3),\\
\dim (n+\med, n-\med, -\med) &=& \frac{1}{6} n(n+1)(n+2)(n+3)(2n+3), \\
\dim (n + \med, m + \med) &=& \frac{2}{3}(n(n+4) - m(m+2) + 3)\\ 
                          & & \times (n+2)(m+1),\nonumber \\
\dim (n-\med, m+ \med) &=& \frac{2}{3} (n(n+4) - m(m+2))(n+1) \\
                       & & \times(m+1). \nonumber
\end{eqnarray}

The spinor components in (\ref{spinor-spin5-1})-(\ref{spinor-spin5-3})
are eigenvectors of $\left(2\mathfrak{C}_2^{\Spin(5)}-\mathfrak{C}_2^{\Spin(6)} \right)$  which appear as a part in the r.h.s of  (\ref{Dirac}):

\begin{eqnarray} \nonumber
\left(2\mathfrak{C}_2^{\Spin(5)}-\mathfrak{C}_2^{\Spin(6)} \right) \Psi^{(n \pm \med, n \pm \med, \med)}_{(n \pm \med, m+\med),\pm} &=&
\left( 2m(m+2)+\frac{5}{4} \right)\Psi^{(n \pm \med, n \pm \med, \med)}_{(n \pm \med, m+\med),\pm}, \\
\left(2\mathfrak{C}_2^{\Spin(5)}-\mathfrak{C}_2^{\Spin(6)} \right) \Psi^{(n+ \med, n- \med , -\med)}_{(n +\med ,m + \med),0} &=& \left( 2n + 2m(m+2) + \frac{13}{4} \right) 
\Psi^{(n+ \med, n- \med , -\med)}_{(n +\med ,m + \med),0}, \nonumber \\ \nonumber 
\left(2\mathfrak{C}_2^{\Spin(5)}-\mathfrak{C}_2^{\Spin(6)}\right) \Psi^{(n+\med, n-\med, -\med)}_{(n -\med ,m + \med),0} &=& \left(-2n+ 2m(m+2)-\frac{11}{4} \right) 
\Psi^{(n+\med, n-\med, -\med)}_{(n -\med ,m + \med),0}. 
\end{eqnarray}

\section{Right-invariant Maurer-Cartan forms}

The $\Spin(6)$ right-invariant Maurer-Cartan forms are defined by $ dR R^{-1}  = -\imath \e_{AB}T_{AB} $, they are dual to the right-invariant vector fields

\be
< \e_{AB}, \calL_{CD}> = \imath {\bf P }_{AB;CD},
\ee
and
\be
{\bf P}_{AB; CD}=\frac{1}{16}\Tr([T_{AB},{\mathcal N}][T_{CD},{\mathcal N}]) .
\ee
By noticing the relations

\begin{eqnarray}
\med\CAS^{\Spin(5)} &=& \calL_{ab} {\bf P}_{ab;cd} \calL_{cd}, \\
\med\CAS^{\Spin(6)} &=& \calL_{ab} (\mathbb{X}_{ab;cd} + 2\mathbb{P}_{ab;cd})  \calL_{cd},
\end{eqnarray}
it follows that the line elements corresponding to these operators are respectively

\be
ds^2_{5} = \e_{ab}{\bf P}_{ab;cd} \e_{cd}, \quad ds^2_{6}  = \e_{ab} \left( \mathbb{X}_{ab;cd} + \frac{\mathbb{P}_{ab;cd}}{2} \right) \e_{cd}.
\ee
From here we obtain the line element (\ref{line-element-spin5}) associated with $\Delta_h$:

\be
d\bar{s}^2 = 4 R^2 \e_{ab} \left( \frac{\mathbb{P}_{ab;cd}}{2} + \frac{\mathbb{X}_{ab;cd}}{1+h}  \right) \e_{cd}.
\ee
In order to fix a normalization for the radii we define the line element (\ref{so6-line-element}) by choosing

\be
ds^2 = \frac{R^2}{4} dn_{AB} dn_{AB}  = 4 R^2 \e_{AB} {\bf P}_{AB;CD} \e_{CD}
\ee
and split it up under $\Spin(5)$ as:

\be
ds^2 = R^2 \left( dn_a^2 + \frac{dn_{ab}^2 - 2dn_a^2}{4} \right) = 4R^2 \e_{ab} \left( \frac{\mathbb{P}_{ab;cd}}{2} + \mathbb{X}_{ab;cd} \right) \e_{cd}.
\ee
We can then read off from (\ref{dbars2}) $R^2_{S^4} = \alpha + 2 \beta$ and $R^2_{S^2} = 4 \beta$. Finally using (\ref{Delta_cas}) we find, as before,

\be
R^2_{S^4} = R^2, \quad \quad R^2_{S^2} = \frac{R^2}{1+h}.
\ee

\end{document}